\documentclass[pre,amsmath,showpacs,preprintnumbers,nofootinbib,preprint]{revtex4}
\usepackage{amsmath,amsfonts,bm}
\usepackage[dvips]{graphicx}

\begin{document}

\title{Path-integral analysis of fluctuation theorems for general Langevin processes}

\author{Vladimir Y. Chernyak $^a$, Michael Chertkov $^{b,c}$, and Christopher Jarzynski $^b$}

\affiliation{$^a$ Department of Chemistry, Wayne State University, 5101 Cass Ave,
Detroit, MI 48202, USA\\
$^b$ T-13 and $^c$ CNLS, Theoretical Division, Los Alamos National
Laboratory, Los Alamos, NM 87545, USA}

\date{\today}

\preprint{LAUR-06-1980}

\begin{abstract}
We examine classical, transient fluctuation theorems within the unifying
framework of Langevin dynamics.
We explicitly distinguish between the effects of non-conservative forces
that violate detailed balance,
and non-autonomous dynamics arising from the variation of an external parameter.
When both these sources of nonequilibrium behavior are present,
there naturally arise two distinct fluctuation theorems.
\end{abstract}

\pacs{47.27.Nz}

\maketitle

\section{Introduction}

The {\it fluctuation theorem} refers to a set of exact
relations describing the statistical mechanics of systems away from
equilibrium, generically expressed by the formula,
\begin{equation}
\frac{p(+\Sigma)}{p(-\Sigma)} = e^{\Sigma}.
\end{equation}
Here $p(\Sigma)$ is the distribution of observed
values of a quantity representing
dissipation or entropy production.
The fluctuation theorem was originally formulated for non-conservative forces
but autonomous dynamics~\cite{93ECM,94ES,95GC,98Kur,99LS},
such as a fluid subject to constant shear,
or a charged particle pushed through a thermal environment
by a constant external field.
Related results have also been derived for
the case of non-autonomous dynamics~\cite{77BoKu,97Jar1,97Jar2,98Cro,99Cro,99Hat,00Cro,01HuSz,01HS},
in which the system of interest is driven away from a stationary state by
the external forcing of a work parameter,
as when a piston is pushed into a gas of particles.
For further extensions and unifying frameworks see
Refs.~\cite{Mae03,MN03,04CCJ,05Sei,05Kur,05RSE,05SpeckSeifert,ImparatoPeliti06},
and for reviews, see Refs.~\cite{02ES,Rit03,PS04,05BLR}.

In the current paper we present an exposition of transient
fluctuation theorems within the path-integral formalism of Langevin dynamics.
We consider two distinct mechanisms for achieving nonequilibrium behavior:
non-conservative forces (explicit violation of detailed balance), and
non-autonomous dynamics (external forcing), and
we distinguish the contributions that each
of these factors makes to the fluctuation theorem.
We will show that when both mechanisms are present,
the definition of $\Sigma$ is not unique, and there arise two
distinct fluctuation theorems.

In the following two Sections, we specify the general class of Langevin dynamics
we consider in this paper, first using the Fokker-Planck formalism
(Section~\ref{sec:model}), then in the path-integral representation
(Section~\ref{sec:pathInt}).
In Section~\ref{sec:ft1} we derive a fluctuation theorem, Eq.~\ref{eq:ftg}, by comparing
the evolution of our system during a {\it forward} process (as an external parameter
is varied from an initial value $A$ to a final value $B$)
to its evolution during the corresponding {\it reverse} process ($B\rightarrow A$).
In Section~\ref{sec:ft2} we obtain a different fluctuation theorem, Eq.~\ref{eq:ftg_dual},
by considering the effect of reversing not only the
protocol for varying the external parameter, but also the underlying dynamics.
In Section~\ref{sec:physics} we discuss physical interpretations of the quantities
appearing in our results, and
in Section~\ref{sec:example} we illustrate these results using a simple model system
with two degrees of freedom.
Finally, in Section~\ref{sec:IFTs} we discuss the {\it integrated} fluctuation
theorems that follow immediately from Eqs.~\ref{eq:ftg} and \ref{eq:ftg_dual};
we present an alternative derivation of these integrated results,
which in turn leads to an extension of the fluctuation theorems derived
in Sections~\ref{sec:ft1} and \ref{sec:ft2}.

\section{Stochastic modeling and Definitions}
\label{sec:model}

Consider an overdamped classical system
described by the stochastic differential equation,
\begin{eqnarray}
\frac{d}{dt} x_i=F_i({\bm x};\lambda)+\xi_i(t;{\bm x};\lambda),
\label{Lang1}
\end{eqnarray}
where $x_i$, with $i=1,\cdots,N$, denote a set of dynamical variables,
and $\lambda$ represents an externally controlled parameter.
${\bm F}$ represents
the deterministic component of the dynamics;
the stochastic component ${\bm \xi}$ is a
$\delta$-correlated noise field, whose mean is zero and whose
pair correlation function is:
\begin{eqnarray}
\Bigl \langle\xi_i(t)
 \, \xi_j(t')\Bigr\rangle=G_{ij}\,\delta(t-t'),
 \label{G}
\end{eqnarray}
where
${\bm G}({\bm x};\lambda)$ is a symmetric, positive definite matrix.
Eq.~\ref{Lang1} can be viewed as the $\Delta\rightarrow 0$
limit of a Markov process with discrete time steps $t_{k+1}-t_k=\Delta$.
In the Appendix, we specify this limiting procedure, and show that
an ensemble of systems governed by Eq.~\ref{Lang1}
is described by a probability density $p({\bm x},t)$, evolving under the
Fokker-Planck equation
\begin{equation}
\label{eq:fp0}
\frac{\partial p}{\partial t} =
-\partial^i ( F_i p) +
\frac{1}{2} \partial^i \Bigl[ G_{ij} (\partial^j p)\Bigr]
\equiv {\cal L}_\lambda p .
\end{equation}
Summation over repeated indices is implied.
As the notation indicates, the Fokker-Planck operator ${\cal L}_\lambda$
depends explicitly on the parameter $\lambda$.
Throughout this paper we will assume that when
$\lambda$ is held fixed, $p$ relaxes exponentially
to a unique stationary distribution
\begin{equation}
\label{eq:pS}
p^S({\bm x};\lambda) = \exp \left[ -\varphi({\bm x};\lambda) \right],
\end{equation}
hence
\begin{equation}
\label{eq:stationaryState}
{\cal L}_\lambda  \exp \left[ -\varphi({\bm x};\lambda) \right] = 0
\end{equation}
and all other eigenvalues of ${\cal L}_\lambda$ are negative and bounded
away from zero.

For fixed $\lambda$, the dynamics are specified by
the vector field ${\bm F}$ and the matrix field ${\bm G}$.
Now let ${\bm\Gamma}({\bm x};\lambda)$ denote the matrix inverse of ${\bm G}({\bm x};\lambda)$,
i.e.\
$\Gamma^{ij}G_{jk} = \delta_k^i$,
and consider two vector fields, ${\bm v}({\bm x};\lambda)$ and ${\bm A}({\bm x};\lambda)$:
\begin{eqnarray}
\label{eq:defv}
v^i &=& 2 \Gamma^{ij} F_j \\
\label{eq:defA}
A^i &=&v^i + \partial^i\varphi \quad.
\end{eqnarray}
These will play an important role in the following analysis.
In terms of ${\bm v}$, Eq.~\ref{eq:fp0} becomes
\begin{equation}
\label{eq:fp1}
\frac{\partial p}{\partial t} =
-\frac{1}{2} \partial^i
\Bigl[ G_{ij} ( v^j - \partial^j ) p \Bigr] =
-\partial^i J_i,
\end{equation}
where ${\bm J}({\bm x},t) = (1/2) {\bm G}({\bm v} - {\bm\nabla}) p$
is the current density.
Substituting $p=p^S = e^{-\varphi}$, we get the stationary current density~\cite{06SpeckSeifert}:
\begin{equation}
\label{eq:JS}
J_i^S({\bm x};\lambda)
= \frac{1}{2} G_{ij}
A^j
e^{-\varphi}
= \left[ F_i + \frac{1}{2} G_{ij} \left( \partial^j \varphi \right) \right] e^{-\varphi}.
\end{equation}
This current is divergenceless:
\begin{equation}
\label{eq:divergenceless}
\partial^i J_i^S = 0.
\end{equation}

If ${\bm v}$ can be written as the
gradient of a scalar field,
$
{\bm v}({\bm x};\lambda) = -{\bm\nabla} U({\bm x};\lambda),
$
then the stationary state is $p^S \propto e^{-U}$,
as seen by inspection of Eq.~\ref{eq:fp1}.
It then follows from Eq.~\ref{eq:defA} that ${\bm A}={\bm 0}$,
which in turn implies a vanishing stationary current,
${\bm J}^S= {\bm 0}$ (Eq.~\ref{eq:JS}).
In this situation we say that the forces acting on the system are
{\it conservative},
or equivalently that detailed balance is satisfied,
and we interpret $p^S \propto e^{-U}$ as an equilibrium (canonical) distribution.
By contrast,
if ${\bm v} \ne - {\bm\nabla} U$, then the forces are
{\it non-conservative}, detailed balance is violated,
and ${\bm A},\,{\bm J}^S \ne {\bm 0}$.
We will thus view ${\bm A}$
as an indicator that distinguishes between
conservative (${\bm A}={\bm 0}$)
and non-conservative (${\bm A} \ne {\bm 0}$) forces.

Another important distinction
is that between {\it autonomous} and {\it non-autonomous} dynamics.
The former refers to the situation in which we observe the evolution
of the system with $\lambda$ held fixed,
while the latter denotes the case when the parameter is varied externally
according to some schedule $\lambda_t$.

Throughout this paper we will consider processes during which
the system evolves over a time interval
$-\tau \le t \le +\tau$,
and we will generally assume that initial conditions are sampled
from the stationary state:
\begin{equation}
\label{eq:stationary_initial}
p({\bm x},-\tau) = \exp \left[ -\varphi({\bm x}; \lambda_{-\tau}) \right] .
\end{equation}
(See however the end of Section~\ref{sec:physics} as well as
Refs.~\cite{04CCJ,05Sei} for discussions of more general initial conditions.)
If the forces are conservative and the dynamics autonomous,
then the system remains in equilibrium over the interval
of observation:
$p({\bm x},t) \propto \exp \left[ -U({\bm x}; \lambda_{\rm fixed}) \right]$.
However, if we have either non-conservative forces
or non-autonomous dynamics, or both,
then we achieve nonequilibrium behavior, for which fluctuation theorems
can be derived.

Finally, rearranging Eq.~\ref{eq:JS} to express ${\bm F}$
as a function of ${\bm G}$, $\varphi$, and ${\bm J}^S$,
we can rewrite the Fokker-Planck operator explicitly in terms
of the stationary density and current:
\begin{equation}
\label{eq:fp2}
{\cal L}_\lambda p=
-\partial^i
\Bigl(
J_i^S  e^{\varphi} p
\Bigr)
- \frac{1}{2}
\partial^i
\Bigl[
G_{ij} e^{-\varphi}
\partial^j\Bigl(
e^{\varphi} p
\Bigr)
\Bigr] .
\end{equation}
By considering various choices of (divergenceless) ${\bm J}^S$, while keeping
${\bm G}$ and $\varphi$ fixed, we explore a family of stochastic dynamics
with a common stationary density but different stationary currents.
If ${\cal L}_\lambda$ corresponds to a given choice of
$\{{\bm J}^S,\varphi,{\bm G}\}$,
then we will use the notation $\hat{\cal L}_\lambda$
to denote the Fokker-Planck operator
corresponding to $\{-{\bm J}^S,\varphi,{\bm G}\}$.
By Eq.~\ref{eq:JS},
the deterministic component of the dynamics associated with
$\hat{\cal L}_\lambda$ is given by
\begin{equation}
\label{eq:Fhat}
\hat F_i = -F_i - G_{ij} \partial^j \varphi.
\end{equation}
Using the path-integral formalism discussed in the following Sections
(Eq.~\ref{eq:Sdiff.dual} in particular)
it is straightforward to establish that
\begin{equation}
\label{eq:dualProperty}
P_\lambda({\bm x}^\prime,\Delta t \vert {\bm x}) \, p^S({\bm x}) =
\hat P_\lambda({\bm x},\Delta t \vert {\bm x}^\prime) \, p^S({\bm x}^\prime)
\end{equation}
for any $\Delta t>0$,
where $P_\lambda$ and $\hat P_\lambda$ are transition probabilities associated
with the dynamics generated by ${\cal L}_\lambda$ and $\hat{\cal L}_\lambda$,
respectively.
Recognizing each side as a joint probability for observing
a pair of events (separated by a time interval $\Delta t$) in the stationary state,
Eq.~\ref{eq:dualProperty} is interpreted as follows:
if we generate an infinitely long trajectory
using the dynamics ${\cal L}_\lambda$,
and we then replace this trajectory by its time-reversed image,
${\bm x}_t \rightarrow {\bm x}_{-t}$, then the new trajectory will be statistically indistinguishable
from a trajectory generated by $\hat{\cal L}_\lambda$.
In particular the two trajectories will be characterized by the same
stationary density, $p^S$, but opposite currents, $\pm{\bm J}^S$.
When two stochastic dynamics are related by Eq.~\ref{eq:dualProperty},
we say that the one is the {\it reversal}~\cite{Norris97},
or the $p^S$-{\it dual}~\cite{Kemeny76}, of the other.
This natural pairing of Fokker-Planck operators $({\cal L}_\lambda, \hat{\cal L}_\lambda)$
will play an important role in Section~\ref{sec:ft2},
where the analysis is very similar to that carried out by Crooks for discrete-time
Markov processes~\cite{00Cro}.
Note that ${\cal L}_\lambda=\hat{\cal L}_\lambda$ when ${\bm J}^S = {\bm 0}$;
in this case Eq.~\ref{eq:dualProperty} is just the familiar statement of
detailed balance associated with conservative forces.
\begin{figure}[htbp]
  \centering
 \includegraphics[scale=0.60]{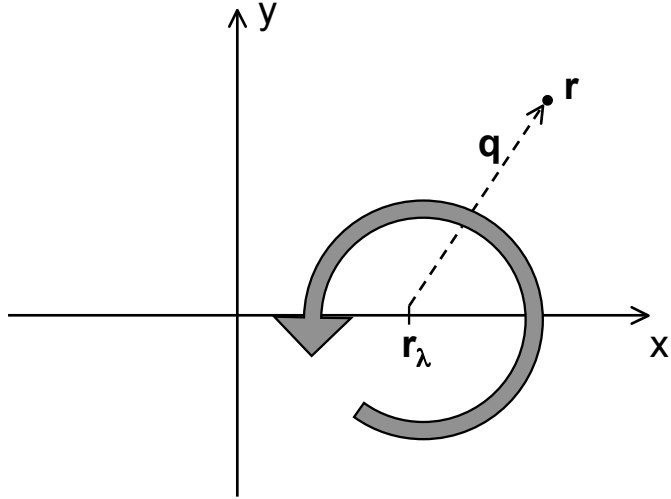}
  \caption{Depiction of the illustrative model described in the text.
The point ${\bm r}$ is the location of a Brownian particle inside a harmonic
well, $kq^2/2$, centered at ${\bm r}_\lambda$.
The grey arrow represents
the angular drift force $f(q) \, \hat{\bm\theta}$, also centered at ${\bm r}_\lambda$.
}
  \label{fig:simpleModel}
\end{figure}

As a simple illustrative model, shown in Fig.~\ref{fig:simpleModel},
consider a particle in two dimensions, ${\bm r}=(x,y)$, with
\begin{equation}
\label{eq:model}
{\bm F}({\bm r};\lambda) =
-\frac{k}{\gamma} {\bm q} +
\frac{1}{\gamma} f(q)
\,\hat{\bm\theta}
\qquad,\qquad
G_{ij}({\bm r};\lambda) = \frac{2}{\gamma} \, \delta_{ij}.
\end{equation}
Here ${\bm q} = {\bm r} - {\bm r}_\lambda$
is the displacement of the particle relative to a point
${\bm r}_\lambda = (\lambda,0)$;
and the unit vector $\hat{\bm\theta}$ is normal
(oriented $90^\circ$ counterclockwise)
to the unit vector $\hat{\bm q} = {\bm q}/q$.
When $f=0$, Eq.~\ref{eq:model} describes a Brownian particle in
a harmonic well centered at ${\bm r}_\lambda$,
with spring constant $k>0$, friction coefficient $\gamma>0$,
and temperature $k_BT=1$.
When $f(q)$ is positive (negative), this particle experiences an additional counter-clockwise
(clockwise) drift around the point ${\bm r}_\lambda$.
For this model, we have:~\cite{non-generic}
\begin{equation}
\label{eq:model.pS}
p^S \propto
e^{-kq^2/2}
\quad,\quad
{\bm A} = f(q) \,\hat{\bm\theta}
\quad,\quad
{\bm J}^S =  \frac{1}{\gamma}\,f(q)
\,\hat{\bm\theta} \, e^{-\varphi}.
\end{equation}
Thus the stationary state is characterized by a Gaussian distribution, $p^S$,
and an average angular drift around the point ${\bm r}_\lambda$.
If ${\cal L}_\lambda$ denotes the Fokker-Planck operator for
a given choice of  $k$, $\gamma$, and $f(q)$,
then $\hat{\cal L}_\lambda$
is obtained by reversing the direction of the angular drift:
\begin{equation}
{\cal L}_\lambda \leftrightarrow\{ k, \gamma, +f(q) \}
\quad \Rightarrow \quad
\hat{\cal L}_\lambda \leftrightarrow \{ k, \gamma, -f(q) \} .
\end{equation}
We will use this model in Section~\ref{sec:example} to illustrate the central
results of this paper.

\section{Path integral formalism}
\label{sec:pathInt}

The main theoretical tool that we will use in this paper is the path-integral
representation of Langevin dynamics~\cite{53OM,wiegel}.
Let $X \equiv \{{\bm x}_t\}_{-\tau}^{+\tau}$
denote a trajectory that specifies the evolution of
the system from $-\tau$ to $+\tau$.
When $\lambda$ is held fixed,
the conditional probability of observing this trajectory, given the
initial microstate ${\bm x}_{-\tau}$, is:
\begin{eqnarray}
\label{eq:condProb.lambdaFixed}
{\cal P}_\lambda\left[ X\vert {\bm x}_{-\tau} \right] &=&
{\cal N} \exp
\left[
-\int_{-\tau}^{+\tau} dt\,{\cal S}_+({\bm x}_t,\dot{\bm x}_t;\lambda)
\right] \\
\label{eq:S}
{\cal S}_+({\bm x},\dot{\bm x};\lambda) &=&
\frac{1}{2}
(\dot x_i - F_i) \Gamma^{ij} (\dot x_i - F_i)
+ \frac{1}{2} \partial^i F_i .
\end{eqnarray}
As discussed in the Appendix, the continuous-time integral
in Eq.~\ref{eq:condProb.lambdaFixed}
is interpreted as the limit of a discrete sum,
using mid-point (Stratonovich) discretization.

Imagine that the system evolves as $\lambda$ is varied externally from
an initial value $\lambda_{-\tau}^{\rm F}=A$ to a final value $\lambda_{+\tau}^{\rm F}=B$,
according to a schedule, or {\it protocol}, $\lambda_t^{\rm F}$.
We refer to this as the {\it forward process}, indicated by the superscript ${\rm F}$.
During this process, the system satisfies
\begin{equation}
\label{eq:eom.F}
\frac{d}{dt} x_i=F_i({\bm x};\lambda_t^{\rm F})+\xi_i(t;{\bm x};\lambda_t^{\rm F}).
\end{equation}
The conditional probability of observing a trajectory $X$ is obtained
by a straightforward generalization of Eq.~\ref{eq:condProb.lambdaFixed}:
\begin{equation}
\label{eq:condProb}
{\cal P}^{\rm F}\left[ X\vert {\bm x}_{-\tau} \right] =
{\cal N} \exp
\left[
-\int_{-\tau}^{+\tau} dt\,{\cal S}_+({\bm x}_t,\dot{\bm x}_t;\lambda_t^{\rm F})
\right] .
\end{equation}
If we sample ${\bm x}_{-\tau}$ from the stationary
distribution $p^S({\bm x};A)$ (Eq.~\ref{eq:stationary_initial}),
the net (unconditional) probability of observing the trajectory $X$ is:
\begin{equation}
\label{eq:fk0}
{\cal P}^{\rm F} [X]
= p_A^S({\bm x}_{-\tau}) \,
{\cal P}^{\rm F}\left[ X\vert {\bm x}_{-\tau} \right],
\end{equation}
where $p_A^S({\bm x}_{-\tau}) \equiv p^S({\bm x}_{-\tau};A)$.

Along with the forward process, we will consider a {\it reverse process},
during which the parameter is manipulated from $B$ to $A$.
Specifically, during the reverse process we have
\begin{equation}
\label{eq:eom.R}
\frac{d}{dt} x_i=F_i({\bm x};\lambda_t^{\rm R})+\xi_i(t;{\bm x};\lambda_t^{\rm R}),
\end{equation}
where
\begin{equation}
\label{eq:reverseProtocol}
\lambda_t^{\rm R} = \lambda_{-t}^{\rm F} .
\end{equation}
The conditional and unconditional probabilities of observing a trajectory $X$
during this process are given by the analogues of Eq.~\ref{eq:condProb} and \ref{eq:fk0}:
\begin{eqnarray}
\label{eq:condProbRev}
{\cal P}^{\rm R}\left[ X\vert {\bm x}_{-\tau} \right] &=&
{\cal N} \exp
\left[
-\int_{-\tau}^{+\tau} dt\,{\cal S}_+({\bm x}_t,\dot{\bm x}_t;\lambda_t^{\rm R})
\right] \\
{\cal P}^{\rm R} [X]
&=& p_B^S({\bm x}_{-\tau}) \,
{\cal P}^{\rm R}\left[ X\vert {\bm x}_{-\tau} \right].
\end{eqnarray}
Here we have assumed the same underlying stochastic dynamics
-- that is, the same family of Fokker-Planck operators ${\cal L}_\lambda$ --
for the reverse process as for the forward process;
the only distinction between the two processes is the protocol for
varying $\lambda$.

\section{Fluctuation theorem for reversed protocol}
\label{sec:ft1}

Now let $X^\dagger \equiv \{{\bm x}_t^\dagger\}_{-\tau}^{+\tau}$
denote the time-reversed ``conjugate twin'' of a trajectory $X$:
\begin{equation}
\label{eq:conjTwin}
{\bm x}_t^\dagger = {\bm x}_{-t},
\end{equation}
and let us compare the probability of observing a trajectory $X$ during
the forward process, with that of its twin
$X^\dagger$ during the reverse process.
Using Eqs.~\ref{eq:reverseProtocol}, \ref{eq:condProbRev} and
\ref{eq:conjTwin}, we get
\begin{equation}
\label{eq:firstStep}
{\cal P}^{\rm R}\left[ X^\dagger\vert {\bm x}_{-\tau}^\dagger \right]  =
{\cal N} \exp
\left[
-\int_{-\tau}^{+\tau} dt\,{\cal S}_+({\bm x}_t^\dagger,\dot{\bm x}_t^\dagger;\lambda_t^{\rm R})
\right] =
{\cal N} \exp
\left[
-\int_{-\tau}^{+\tau} dt\,{\cal S}_-({\bm x}_t,\dot{\bm x}_t;\lambda_t^{\rm F})
\right] ,
\end{equation}
where
\begin{equation}
\label{eq:defSminus}
{\cal S}_-({\bm x},\dot{\bm x};\lambda) \equiv
{\cal S}_+({\bm x},-\dot{\bm x};\lambda).
\end{equation}
The definitions of ${\bm v}$, ${\bm A}$, and ${\cal S}_\pm$
then give us
\begin{equation}
\label{eq:Sdiff}
{\cal S}_+ - {\cal S}_- = -\dot x_j v^j = \dot x_j
\left( \partial^j \varphi - A^j  \right),
\end{equation}
which combines with Eqs.~\ref{eq:condProb} and \ref{eq:firstStep} to yield
\begin{equation}
\label{eq:ratio00}
\frac
{{\cal P}^{\rm F}\left[ X\vert {\bm x}_{-\tau} \right]}
{{\cal P}^{\rm R}[ X^\dagger\vert {\bm x}_{-\tau}^\dagger ]}
=
\exp\left(
\int^{\rm F} dt\, \dot x_j v^j
\right),
\end{equation}
where $\int^{\rm F}$ denotes evaluation along
the forward protocol and trajectory $X$.
When the diffusion tensor is a constant,
${\bm G}({\bm x};\lambda) = 2D\delta_{ij}$,
this ratio is equivalent to a result
derived previously by Seifert (Eq.~14 of Ref.~\cite{05Sei}),
and extended to inertial systems by
Imparato and Peliti (Eq.~34 of Ref.~\cite{ImparatoPeliti06}).
From Eq.~\ref{eq:ratio00}, we obtain
\begin{equation}
\label{eq:ratio0}
\frac
{{\cal P}^{\rm F}[X]}{{\cal P}^{\rm R}[X^\dagger]}
=
\frac
{  p_A^S({\bm x}_{-\tau})  {\cal P}^{\rm F}[X \vert {\bm x}_{-\tau}]}
{  p_B^S({\bm x}_{-\tau}^\dagger)  {\cal P}^{\rm R}[X^\dagger \vert {\bm x}_{-\tau}^\dagger]}
=
\exp\left(
\Delta\varphi +
\int^{\rm F} dt\, \dot x_j v^j
\right),
\end{equation}
where
\begin{equation}
\label{eq:Delta_phi}
\Delta\varphi \equiv
\varphi({\bm x}_\tau;B) - \varphi({\bm x}_{-\tau};A)
=
\int^{\rm F} dt\,\left(
\dot\lambda_t^{\rm F} \, \frac{\partial\varphi}{\partial\lambda} +
\dot x_j \, \partial^j\varphi
\right).
\end{equation}

Let us now rewrite Eq.~\ref{eq:ratio0} so that the quantity in the
exponent is manifestly a sum of contributions
representing non-autonomous dynamics and non-conservative forces.
Defining
\begin{equation}
\label{eq:defYF}
Y^{\rm F} \equiv \int^{\rm F} dt\, \dot\lambda_t^{\rm F} \, \frac{\partial\varphi}{\partial\lambda} ,
\end{equation}
we obtain
\begin{equation}
\label{eq:ratio}
\frac
{{\cal P}^{\rm F}[X]}{{\cal P}^{\rm R}[X^\dagger]}
=
\exp\left(
Y^{\rm F} + \int^{\rm F} d{\bm x}\cdot {\bm A} \right).
\end{equation}
A non-zero value of $Y^{\rm F}$ is a signature of non-autonomous dynamics
(Eq.~\ref{eq:defYF}), whereas ${\bm A} \ne 0$ indicates
non-conservative forces (Section~\ref{sec:model}).
Thus we associate the two terms, $Y^{\rm F}$ and $\int^{\rm F} d{\bm x}\cdot {\bm A}$,
with the two mechanisms for achieving
nonequilibrium behavior identified at the end of Section \ref{sec:model}.
Let $R$ denote the sum of these two terms:
\begin{equation}
\label{eq:Rdef}
R^{\rm F}[X] \equiv Y^{\rm F} + \int^{\rm F} d{\bm x}\cdot{\bm A}.
\end{equation}
When the dynamics are autonomous and the forces conservative,
both terms are equal to zero, hence ${\cal P}[X] = {\cal P}[X^\dagger]$:
in equilibrium, any sequence of events is as likely as the reverse sequence;
see e.g.\ the discussion following Eq.~[13] of Ref.~\cite{99BDKA}.

For the reverse process we similarly define
\begin{equation}
\label{eq:defYR}
Y^{\rm R} \equiv \int^{\rm R} dt\, \dot\lambda_t^{\rm R} \, \frac{\partial\varphi}{\partial\lambda}
\quad,\quad
R^{\rm R} \equiv Y^{\rm R} + \int^{\rm R} d{\bm x}\cdot{\bm A}.
\end{equation}
The quantities $Y$ and $R$ are odd under time-reversal,
in the following sense:
\begin{equation}
\label{eq:odd}
Y^{\rm R}[X^\dagger] = -Y^{\rm F}[X]
\qquad,\qquad
R^{\rm R}[X^\dagger] = -R^{\rm F}[X].
\end{equation}

With this formalism in place, we now derive a fluctuation theorem for $R = Y + \int d{\bm x}\cdot{\bm A}$.
Let $\rho^{\rm F}(R)$ denote the distribution of $R$ values for
an ensemble of realizations of the forward process,
and define $\rho^{\rm R}(R)$ analogously for the reverse process.
Then
\begin{eqnarray}
\rho^{\rm F}(R)
&=&
\int {\cal D} X \, {\cal P}^{\rm F}[X] \,
\delta\left(R - R^{\rm F}[X]\right) \\
&=&
\int {\cal D} X \, {\cal P}^{\rm R}[X^\dagger] \,
\exp\left( R^{\rm F}[X]\right) \,
\delta\left(R - R^{\rm F}[X]\right) \\
&=&
\exp(R) \,
\int {\cal D} X^\dagger \,
{\cal P}^{\rm R}[X^\dagger] \,
\delta\left(R + R^{\rm R}[X^\dagger]\right) ,
\end{eqnarray}
where ${\cal D}X = \prod_{k=0}^K d^Nx_k$ specifies an integral
over all possible trajectories $X$ (see Appendix).
We have used Eq.~\ref{eq:ratio} to get from the first line to the second,
and Eq.~\ref{eq:odd} to get to the third.
Note also the change of variables, ${\cal D} X \rightarrow {\cal D} X^\dagger$.
Recognizing the final integral as $\rho^{\rm R}(-R)$, we obtain the desired
fluctuation theorem~\cite{05Sei}:
\begin{equation}
\label{eq:ftg}
\frac{\rho^{\rm F}(+R)}{\rho^{\rm R}(-R)} = \exp (R).
\end{equation}

\section{Fluctuation theorem for reversed protocol and dynamics}
\label{sec:ft2}

The evolution of the system is influenced by both the protocol for varying $\lambda$,
and the stochastic dynamics that define the Langevin transition rates.
In the previous Section we assumed that the forward and
reverse processes are defined by conjugate protocols
but the same underlying dynamics
(Eqs.~\ref{eq:eom.F}, \ref{eq:eom.R}).
Thus the reverse process was defined relative to the forward process
by the replacement
\begin{equation}
\label{eq:revProtocolOnly}
\{ {\cal L}_\lambda , \lambda_t^{\rm F} \}
\rightarrow
\{ {\cal L}_\lambda , \lambda_t^{\rm R} \}.
\end{equation}
In this Section we obtain a different fluctuation theorem
by imagining that the reverse process is characterized by a reversal of
both the protocol and the underyling dynamics.

Specifically, we imagine that during the forward process the system satisfies
Eq.~\ref{eq:eom.F}, as in the previous Section.
However, for the reverse process, we take
\begin{eqnarray}
\label{eq:dual.dxdt}
\frac{d}{dt} x_i=\hat F_i({\bm x};\lambda_t^{\rm R})+\xi_i(t;{\bm x};\lambda_t^{\rm R}),
\end{eqnarray}
where $\hat{\bm F} = -{\bm F} - {\bm G\,\nabla}\varphi$ (Eq.~\ref{eq:Fhat}),
rather than Eq.~\ref{eq:eom.R}.
Thus the reverse process is now defined by the replacement
\begin{equation}
\{ {\cal L}_\lambda , \lambda_t^{\rm F} \}
\rightarrow
\{ \hat{\cal L}_\lambda , \lambda_t^{\rm R} \} .
\end{equation}
In this situation the
conditional probability of a trajectory $X$ during the reverse process is
\begin{equation}
\label{eq:condProbRev.dual}
\hat{\cal P}^{\rm R}\left[ X\vert {\bm x}_{-\tau} \right] =
{\cal N} \exp
\left[
-\int_{-\tau}^{+\tau} dt\,\hat{\cal S}_+({\bm x}_t,\dot{\bm x}_t;\lambda_t^{\rm R})
\right] ,
\end{equation}
where $\hat{\cal S}_+$ is defined as its counterpart ${\cal S}_+$, but with
${\bm F}$ replaced by $\hat{\bm F}$.
Let us similarly define
$\hat{\cal S}_-({\bm x},\dot{\bm x};\lambda) \equiv
\hat{\cal S}_+({\bm x},-\dot{\bm x};\lambda)$
(as in Eq.~\ref{eq:defSminus}).
By direct evaluation we obtain
\begin{equation}
\label{eq:Sdiff.dual}
{\cal S}_+ - \hat{\cal S}_- =
\dot x_i \partial^i\varphi +
e^{\varphi} \partial^i
\left[
\left(
F_i - \frac{1}{2} G_{ij} \partial^j
\right)
e^{-\varphi}
\right]
=
\dot x_i \partial^i\varphi,
\end{equation}
since the quantity in square brackets is just the $i$'th component of
the divergenceless stationary current (Eq.~\ref{eq:JS}).

Repeating the steps of Section~\ref{sec:ft1},
but with Eq.~\ref{eq:Sdiff.dual} in place of Eq.~\ref{eq:Sdiff}, we get
\begin{equation}
\label{eq:ratio.dual0}
\frac
{{\cal P}^{\rm F}\left[ X\vert {\bm x}_{-\tau} \right]}
{\hat{\cal P}^{\rm R}[ X^\dagger\vert {\bm x}_{-\tau}^\dagger ]}
=
\exp\left(
-\int^{\rm F} dt\, \dot x_j \partial^i\varphi
\right),
\end{equation}
and in turn
\begin{equation}
\label{eq:ratio.dual}
\frac
{{\cal P}^{\rm F}[X]}{\hat{\cal P}^{\rm R}[X^\dagger]}
=
\exp\left(
\Delta\varphi -
\int^{\rm F} dt\, \dot x_i \partial^i\varphi
\right)
=
\exp\left(
\int^{\rm F} dt\, \dot\lambda_t^{\rm F} \, \frac{\partial\varphi}{\partial\lambda}
\right)
=
\exp \left( Y^{\rm F} \right) .
\end{equation}
This is identical to Eq.~\ref{eq:ratio}, except that the term
$\int^{\rm F} d{\bm x}\cdot {\bm A}$
no longer appears inside the exponent on the right side.
In effect, by using $\hat{\cal L}_\lambda$ for the reverse process,
Eq.~\ref{eq:dual.dxdt}, we have ``gauged away'' the non-conservative contribution
arising from ${\bm A}$
(compare Eqs.~\ref{eq:Sdiff} and \ref{eq:Sdiff.dual}),
leaving only the non-autonomous contribution associated with
the variation of $\lambda$.

Eq.~\ref{eq:ratio.dual} leads to the analogue of Eq.~\ref{eq:ftg}:
\begin{equation}
\label{eq:ftg_dual}
\frac{\rho^{\rm F}(+Y)}{\hat\rho^{\rm R}(-Y)} =  \, \exp(Y) ,
\end{equation}
where $\rho^{\rm F}$
is the distribution of $Y$ values for the forward process
(${\cal L}_\lambda,\lambda_t^{\rm F}$),
and $\hat\rho^{\rm R}$ is defined similarly for the reverse process
($\hat{\cal L}_\lambda,\lambda_t^{\rm R}$).
This is a continuous-time analogue of
the fluctuation theorem obtained by Crooks for discrete-time processes
(our Eqs.~\ref{eq:ratio.dual0} and \ref{eq:ftg_dual} correspond
to Eqs.~13 and 20 of Ref.~\cite{00Cro}.)

We thus see that two different fluctuation theorems (Eqs.~\ref{eq:ftg}, \ref{eq:ftg_dual})
emerge naturally in the Langevin formalism that we have considered.
The relationship between these two results can be appreciated by separately
considering the three different combinations for achieving  nonequilibrium behavior:
\begin{itemize}
\item{\it Non-conservative forces, autonomous dynamics} (${\bm A} \ne {\bm 0}$, $\dot\lambda = 0$).
In this case $Y=0$ by definition,
hence Eq.~\ref{eq:ftg_dual} has no meaning.
Thus we get a unique fluctuation theorem,
Eq.~\ref{eq:ftg}, for a quantity $R = \int d{\bm x}\cdot{\bm A}$ associated with the violation
of detailed balance.
\item{\it Conservative forces, non-autonomous dynamics} (${\bm A} = {\bm 0}$, $\dot\lambda\ne 0$).
In this case ${\cal L}_\lambda = \hat{\cal L}_\lambda$
and $Y=R$.
Thus Eq.~\ref{eq:ftg} is equivalent to Eq.~\ref{eq:ftg_dual},
so we again obtain a unique fluctuation theorem, this time for the quantity
$Y = R = \int dt\,\dot\lambda\,\partial\varphi/\partial\lambda$
associated with external forcing.
\item{\it Non-conservative forces, non-autonomous dynamics} (${\bm A} \ne {\bm 0}$, $\dot\lambda \ne 0$).
Eqs.~\ref{eq:ftg} and \ref{eq:ftg_dual} now represent two different, but equally
valid, fluctuation theorems,
for the distinct quantities
$Y = \int dt\,\dot\lambda\,\partial\varphi/\partial\lambda$ and
$R = Y +  \int d{\bm x}\cdot{\bm A}$.
\end{itemize}

\section{Physical interpretations}
\label{sec:physics}

To this point our analysis has been mostly abstract and mathematical.
For the Langevin process defined by Eq.~\ref{Lang1}, we have derived
two fluctuation theorems, and in Section~\ref{sec:IFTs} below, closely related {\it integrated}
fluctuation theorems are obtained.
In the present Section we briefly discuss physical interpretations of the quantities
appearing in these results.

Eq.~\ref{Lang1} can be used to model the microscopic evolution of a system in contact with
a thermal reservoir at temperature $T$, in the overdamped limit, with $\lambda$ playing
the role of an externally manipulated work parameter.
The stationary state is then given by the Boltzmann-Gibbs distribution
$p^S = e^{\beta(F-U)}$, with $\beta = (k_B T)^{-1}$, hence
\begin{equation}
\varphi({\bm x};\lambda) = \beta U({\bm x};\lambda) - \beta F(\lambda),
\end{equation}
where $U({\bm x};\lambda)$ represents the internal energy of the system,
and $F(\lambda)$ is the parameter-dependent free energy.
(We leave implicit the temperature dependence of $F$.)
We generically expect detailed balance to hold in such an equilibrium state,
hence the dynamics are conservative: ${\bm A} , \, {\bm J}^S=0$.
Under these circumstances, we get
\begin{equation}
Y = R =  \beta \left(
\int dt\, \dot\lambda \frac{\partial U}{\partial\lambda} - \Delta F \right)
= \beta W_{\rm diss},
\end{equation}
where $\Delta F \equiv F(B)-F(A)$,
and $W_{\rm diss} = W - \Delta F$ is physically interpreted as {\it dissipated work}~\cite{97Jar1}.
There is no distinction between Eqs.~\ref{eq:ftg} and \ref{eq:ftg_dual} in this situation;
both are equivalent to the Crooks fluctuation theorem~\cite{99Cro},
and the corresponding integrated result (Eq.~\ref{eq:IFTs} below)
is the nonequilibrium work theorem,
\begin{equation}
\left\langle e^{-\beta W} \right\rangle = e^{-\beta\Delta F}.
\end{equation}

When ${\bm A} \ne {\bm 0}$, the dynamics are non-conservative.
To gain insight into the physical meaning of the vector field ${\bm A}$,
consider an overdamped, one-dimensional
Brownian particle at temperature $T$, with periodic boundary conditions
(see e.g.\ Figure 1 of Ref.~\cite{05SpeckSeifert}):
\begin{equation}
\label{eq:periodicParticle}
\gamma \dot x = - \frac{\partial U}{\partial x} + f + \zeta(t),
\end{equation}
where $\gamma>0$ and $f$ are constants;
$U(x;\lambda)$ is a periodic potential;
and the noise term satisfies
$\langle \zeta(0) \zeta(t) \rangle = 2\gamma k_B T \delta(t)$.
When $\lambda$ is held fixed the system relaxes to $p^S = e^{-\varphi}$.
Comparing with Eq.~\ref{Lang1},
and using the definition of ${\bm A}$ (which for this example is a scalar field),
we obtain
\begin{equation}
A = \beta \left( -U^\prime + f \right) + \varphi^\prime,
\end{equation}
where the primes denote $\partial/\partial x$.
In previous studies of this example by Hatano and Sasa~\cite{01HS}
and Speck and Seifert~\cite{05SpeckSeifert},
the quantity $Q_{\rm hk} \equiv \beta^{-1} \int dx\,A$ was identified as the
``house-keeping heat'', a concept introduced earlier by
Oono and Paniconi~\cite{98OP}.
In the stationary state,
$Q_{\rm hk}$ represents the heat absorbed by the external reservoir;
this fluctuating quantity grows linearly with time, on average,
and can be viewed as the thermodynamic price that must continually
be paid to maintain the system away from equilibrium.
Equivalently, $\beta Q_{\rm hk} = \int dx \, A$ is the increase in the entropy
of the reservoir.
We speculate that this interpretation remains valid more generally:
when Eq.~\ref{Lang1} models a system in contact with thermal
surroundings, perhaps including multiple heat baths,
$\int d{\bm x}\cdot{\bm A}$ is the
entropy generated in these surroundings, in the stationary state.
This interpretation emphasizes the physical connection between
entropy generation and the violation of detailed balance.

We see that $Y$ can be interpreted as dissipated work (in units of $k_BT$)
when the dynamics are conservative,
and $\int d{\bm x}\cdot {\bm A}$ as the entropy generation needed to
maintain a nonequilibrium stationary state.
For the non-autonomous, non-conservative case,
we do not have simple thermodynamic interpretations of $Y$ and $R$.
However,
we can provide some intuition regarding the difference between these quantities
by considering a quasi-static process, with $\lambda$ varied slowly from $A$ to $B$.
Then $Y \rightarrow 0$~\cite{01HS}, whereas $\int d{\bm x}\cdot{\bm A}$ can be
expected to grow diffusively, with slowly changing drift and diffusion constants.
Hence as $\tau\rightarrow\infty$, the distribution of $Y$ values tends to a delta-function,
while the mean and variance of the distribution of $R$ values scale like $\tau$.

Finally, let us briefly consider a generalization of Eq.~\ref{eq:ftg}.
The quantity $\Delta\varphi$ appearing in Eq.~\ref{eq:ratio0}
is essentially a boundary term~\cite{05Sei},
arising from the ratio of the probabilities of sampling the initial conditions
of the conjugate pair of trajectories, $X$ and $X^\dagger$.
If we choose to sample initial conditions from distributions
other than the stationary distribution, $p^S$, then this term must correspondingly
be modified.
Specifically, consider a family of normalized distributions
$q_\lambda({\bm x}) = \exp[-\psi({\bm x};\lambda)]$,
where $\psi({\bm x};\lambda)$ is arbitrary (apart from the normalization condition).
Now imagine that we define our forward (reverse) process by sampling the initial
conditions from $q_A$ ($q_B$) rather than $p_A^S$ ($p_B^S$).
The boundary term in Eq.~\ref{eq:ratio0} then changes from
$\Delta\varphi$ to $\Delta\psi$, which
ultimately leads to the result
\begin{equation}
\label{eq:ftg.psi}
\frac{\rho^{\rm F}(+R^\psi)}{\rho^{\rm R}(-R^\psi)} = \exp (R^\psi),
\end{equation}
where
\begin{equation}
R^{\psi,{\rm F}} =
\int^{\rm F} dt\,\dot\lambda_t^{\rm F}\,\frac{\partial\psi}{\partial\lambda} +
\int^{\rm F} d{\bm x}\cdot{\bm A}^\psi
\quad,\quad
{\bm A}^\psi = {\bm v} + {\bm\nabla}\psi,
\end{equation}
and similarly for the reverse process.
Eq.~\ref{eq:ftg.psi} thus generalizes Eq.~\ref{eq:ftg} to allow for
initial conditions sampled from arbitrary distributions.

As a specific example,
suppose there exists a natural decomposition
$\beta {\bm v} = -{\bm\nabla} U + {\bm f}$,
where ${\bm f}$ cannot be expressed as the gradient
of a scalar field, e.g.\ imagine a Brownian particle exploring
a potential landscape $U({\bm x};\lambda)$,
but also subject to a non-conservative force ${\bm f}({\bm x};\lambda)$.
This is a generalization of the one-dimensional example
discussed above (Eq.~\ref{eq:periodicParticle}).
Now suppose we sample initial conditions from the
canonical distribution $q \propto \exp (-\beta U)$, rather than the stationary distribution
$p^S = \exp (-\varphi)$.
This corresponds to the choice $\psi = \beta U$,
where for simplicity we have incorporated the free energy $F(\lambda)$
into the definition of $U({\bm x};\lambda)$.
The quantity $R^\psi$ appearing in Eq.~\ref{eq:ftg.psi} then becomes
\begin{equation}
\label{eq:R_psi}
R^{\psi,{\rm F}} =
 \beta \int^{\rm F} dt\, \dot\lambda_t^{\rm F} \, \frac{\partial U}{\partial\lambda}
 +
\beta  \int^{\rm F} d{\bm x}\cdot{\bm f}.
 \end{equation}
We thus get a fluctuation theorem for a quantity
$R^\psi$ that is physically interpreted as the sum of contributions due
the external variation of a conservative potential, $U$,
and the path integral of a non-conservative force, ${\bm f}$.
Neither term depends on the stationary distribution $\varphi$.
This result was originally obtained by Kurchan,
first for autonomous dynamics~\cite{98Kur}
($R^\psi = \beta \int d{\bm x}\cdot{\bm f}$),
and more recently for non-autonomous dynamics~\cite{05Kur},
assuming a spatially independent diffusion coefficient.

\section{Illustrative example}
\label{sec:example}

In this Section we illustrate the fluctuation theorems derived above,
using the example introduced at the end of Section~\ref{sec:model}.
We begin with the autonomous case.
If $f(q)=0$, then we simply have a Brownian particle fluctuating in
equilibrium in a two-dimensional harmonic well.
When $f(q)>0$, the particle is subject to an additional
angular drift, oriented counter-clockwise.
Let us picture this drift to be the result of a vortex in the surrounding
thermal medium, centered at ${\bm r}_\lambda$.
In this case the integral
\begin{equation}
\label{eq:circDef}
\int d{\bm x}\cdot{\bm A}
= \int dt\, \dot{\bm r}_t \cdot \hat{\bm\theta}  \, f({\bm q}_t)
\equiv {\cal C}[X]
\end{equation}
provides a measure of the counterclockwise motion of the particle.
We will refer to ${\cal C}$ as the {\it circulation} associated with a given
trajectory $X$.
Since we are considering autonomous dynamics,
we have $R = {\cal C}$ (see Eq.~\ref{eq:Rdef}),
and Eq.~\ref{eq:ftg} becomes
\begin{equation}
\label{eq:ftg.modelA}
\frac{\rho(+{\cal C})}{\rho(-{\cal C})} = \exp ({\cal C}).
\end{equation}
Here $\rho({\cal C})$ is the probability distribution of observing a circulation
${\cal C}$, over the interval of observation,
assuming initial conditions sampled from the stationary distribution (Eq.~\ref{eq:model.pS}).
Since $\lambda$ is fixed,
there is no distinction between the forward and reverse process.
Eq.~\ref{eq:ftg.modelA} implies that positive values of circulation are more likely than
negative values, as expected:
the particle tends to flow with, rather than against, the vortex.

Now consider the case of non-autonomous dynamics, but conservative forces,
$f(q)=0$.
For specificity, imagine that during the forward process we vary $\lambda$
at a constant rate from $A$ to $B$, e.g.~we move the point ${\bm r}_\lambda$
rightward along the ${\bm x}$-axis.
Thus we drag the Brownian particle through a
thermal medium, by moving the harmonic potential in which it is trapped.
During the reverse process we go from $B$ back to $A$, at the same speed.
Since the forces are conservative
(${\bm A},\,{\bm J}^S=0$, hence ${\cal L}_\lambda=\hat{\cal L}_\lambda$),
there is no distinction between the predictions of Section~\ref{sec:ft1}
and those of Section~\ref{sec:ft2};
moreover there is no tendency for the particle to circulate in one
direction or the other.
We have
\begin{equation}
\label{eq:work.example}
R = Y \equiv \int dt\,\dot\lambda\,\frac{\partial\varphi}{\partial\lambda}
= -k \int d\lambda \, (x_t - \lambda_t),
\end{equation}
where $x_t$ is the $x$-component of the particle's position at time $t$.
Physically,
$-k(x-\lambda)\,d\lambda$ is the incremental work
required to displace the potential well by an amount $d\lambda$,
hence $Y$ is equal to the total external work performed
during a realization of either the forward or the reverse process
(in units of temperature, since $k_BT=1$ for this example).
Eqs.~\ref{eq:ftg} and \ref{eq:ftg_dual} are identical in this situation,
and are expressed as
\begin{equation}
\label{eq:ftg.modelB}
\frac{\rho^{\rm F}(+Y)}{\rho^{\rm R}(-Y)} = \exp (Y).
\end{equation}
Positive values of work are more likely than negative values,
in agreement with the second law of thermodynamics.

Having obtained one prediction for the case of autonomous dynamics
and non-conservative forces (Eq.~\ref{eq:ftg.modelA} for ${\cal C}$),
and another for conservative forces
and non-autonomous dynamics (Eq.~\ref{eq:ftg.modelB} for $Y$),
we now consider the combination of non-autonomous dynamics
and non-conservative forces.
Thus ${\bm r}_\lambda$ now specifies
the moving center of both a harmonic well and a vortex in the thermal medium.
Let ${\cal L}_\lambda$ denote the stochastic dynamics for a given
choice $f(q)>0$, describing a counter-clockwise vortex;
then $\hat{\cal L}_\lambda$ corresponds to the replacement
$f \rightarrow -f$, describing a clockwise vortex.

If we use ${\cal L}_\lambda$ for both the forward and the reverse
processes (Eqs.~\ref{eq:eom.F}, \ref{eq:eom.R}),
then this amounts to dragging the particle either
rightward ($A\rightarrow B$) or leftward ($B\rightarrow A$), but
in either case the particle tends to circulate counter-clockwise
around the moving point ${\bm r}_\lambda$.
The fluctuation theorem in this situation,
Eq.~\ref{eq:ftg}, involves the quantity $R = Y + {\cal C}$,
with external work $Y$ and circulation ${\cal C}$ as defined
by Eqs.~\ref{eq:work.example} and \ref{eq:circDef} above.
On the other hand, if we use $\hat{\cal L}_\lambda$ for the reverse
process (Eq.~\ref{eq:dual.dxdt}), then the particle tends to circulate counter-clockwise  as it is
dragged rightward during the forward process,
but clockwise as it is dragged leftward during the reverse process.
In this case the applicable fluctuation theorem is Eq.~\ref{eq:ftg_dual}
for the work $Y$;
the circulation ${\cal C}$ no longer contributes to the quantity
appearing in the fluctuation theorem.

\section{Integrated fluctuation theorems and further extensions}
\label{sec:IFTs}

Multiplying both sides of Eq.~\ref{eq:ftg} by $\rho^{\rm R}(-R) \exp(-R)$,
then integrating with respect to $R$, and performing similar manipulations
on Eq.~\ref{eq:ftg_dual}, we are led to the {\it integrated} results,
\begin{equation}
\label{eq:IFTs}
\left\langle \exp(-R) \right\rangle^{\rm F} = 1
\quad,\quad
\left\langle \exp(-Y) \right\rangle^{\rm F} = 1,
\end{equation}
where the angular brackets denote an average over realizations of the forward process.
For the general case of non-autonomous and non-conservative dynamics,
with initial conditions sampled from the stationary state,
these results have been discussed by Hatano and Sasa~\cite{01HS} (for $Y$),
and by Speck and Seifert~\cite{05SpeckSeifert} (for $R$).
Unlike Eqs.~\ref{eq:ftg} and \ref{eq:ftg_dual},
the integrated fluctuation theorems
can be stated without any reference to the reverse process,
suggesting that they can be derived by some means other than a comparison
between forward and reverse processes.
We now sketch such a derivation,
which in turn leads to a further generalization of both integrated and non-integrated
results (Eqs.~\ref{eq:IFT.gen}, \ref{eq:ftg.gen}).

We assume our system obeys the equation of motion
\begin{equation}
\label{eq:eom}
\frac{d}{dt} x_i=F_i({\bm x};\lambda_t)+\xi_i(t;{\bm x};\lambda_t),
\end{equation}
where $\lambda_t$ denotes a protocol for varying the parameter $\lambda$.
For a trajectory generated during this process,
consider the variable
\begin{equation}
\label{eq:defomega}
\omega_t =
\int_{-\tau}^t dt^\prime \,
\left[
\dot\lambda \frac{\partial\varphi}{\partial\lambda} +
\alpha\, \dot{\bm x}\cdot {\bm A} +
\frac{1}{2}\alpha(\alpha-1)
G_{ij} A^i A^j
\right] ,
\end{equation}
where $\alpha$ is a constant,
and the integrand is evaluated along the trajectory.
For $\alpha=0$ and $1$, we get
$\omega_\tau = W$ and $R$, respectively.

Now consider a function
\begin{equation}
{\cal G}({\bm x},t) =
\Bigl\langle \delta( {\bm x} - {\bm x}_t ) \, \exp(-\omega_t) \Bigr\rangle,
\end{equation}
where the average is taken over an ensemble of realizations of the process,
and ${\bm x}_t$ is the microstate at time $t$ during a given realization
This function can be viewed as a ``weighted'' phase space
density, in which each realization carries a time-dependent statistical weight
$\exp(-\omega_t)$;
similar constructions have been considered in
Refs.~\cite{97Jar2,01HuSz,04CCJ,ImparatoPeliti06}.
Following a procedure analogous to the derivation of
Eq.~\ref{eq:fp0} in the Appendix, we obtain an evolution equation
for this density:
\begin{equation}
\label{eq:eom.calG}
\frac{\partial}{\partial t} {\cal G} =
{\cal L}_\lambda {\cal G} -
\dot{\lambda}\frac{\partial\varphi}{\partial\lambda} {\cal G}  +
\alpha \partial^i \left( G_{ij} A^j {\cal G} \right) .
\end{equation}
The first term on the right represents the dynamical component
of the evolution (Eq.~\ref{eq:fp0}),
while the other two arise from the weight $\exp(-\omega_t)$
assigned to each trajectory in the ensemble.
Using Eqs.~\ref{eq:stationaryState}, \ref{eq:JS}, and \ref{eq:divergenceless},
we see by inspection that the {\it ansatz}
${\cal G}({\bm x},t) = \exp [ -\varphi({\bm x};\lambda_t) ]$
solves Eq.~\ref{eq:eom.calG}.
Thus when initial conditions are sampled from the stationary
distribution $p^S({\bm x};\lambda_{-\tau})$, we have
\begin{equation}
\label{eq:Gg}
\Bigl\langle \delta( {\bm x} - {\bm x}_t) \, \exp(-\omega_t) \Bigr\rangle
= \exp \left[ -\varphi({\bm x};\lambda_t) \right] .
\end{equation}
Setting $t=+\tau$ and integrating both sides with respect to ${\bm x}$, we get
\begin{equation}
\label{eq:IFT.gen}
\left\langle \exp(-\omega_\tau) \right\rangle = 1 .
\end{equation}
This represents a family of predictions -- parametrized by the
value of $\alpha$ -- that are all satisfied for the same process,
Eq.~\ref{eq:eom}.
For the choices $\alpha=1, 0$, Eq.~\ref{eq:IFT.gen} reduces to Eq.\ref{eq:IFTs}.

Since Eq.~\ref{eq:IFT.gen} generalizes Eq.~\ref{eq:IFTs} to arbitrary $\alpha$,
it is natural to search for a corresponding extension
of Eqs.~\ref{eq:ftg} and \ref{eq:ftg_dual}.
Such a result indeed exists, and is given by:
\begin{equation}
\label{eq:ftg.gen}
\frac{\rho^{\rm F}(+\omega_\tau)}{\rho_\alpha^{\rm R}(-\omega_\tau)} =  \, \exp(\omega_\tau) .
\end{equation}
Here $\rho^{\rm F}$ and $\rho_\alpha^{\rm R}$ denote ensemble distributions of $\omega_\tau$
for a forward and a reverse process.
The forward process is defined as earlier,
for the protocol $\lambda_t^{\rm F}$ and a family of Fokker-Planck operators
${\cal L}_\lambda \leftrightarrow \{ {\bm J}^S , \varphi , {\bm G} \}$.
The reverse process uses $\lambda_t^{\rm R}$ and
\begin{equation}
\label{eq:Lrev.gen}
{\cal L}_\lambda^{\alpha} \leftrightarrow
\{ {\bm J}_\alpha^S , \varphi , {\bm G} \} ,
\end{equation}
where ${\bm J}_\alpha^S \equiv (2\alpha -1) {\bm J}^S$.
The derivation of Eq.~\ref{eq:ftg.gen} follows the same steps as in Sections~\ref{sec:ft1}
and \ref{sec:ft2}, starting from a generalization
of Eqs.~\ref{eq:Sdiff} and \ref{eq:Sdiff.dual},
\begin{equation}
{\cal S}_+ - {\cal S}_-^{\alpha} =
\dot x_i
\left( \partial^i \varphi - \alpha A^i \right)
- \frac{1}{2}\alpha(\alpha-1) G_{ij}A^i A^j ,
\end{equation}
with ${\cal S}_-^{\alpha}$ defined for ${\cal L}_\lambda^{\alpha}$,
as ${\cal S}_-$ and $\hat{\cal S}_-$ were defined for ${\cal L}_\lambda$
and $\hat{\cal L}_\lambda$.
For $\alpha=1, 0$ we get ${\bm J}_\alpha^S = \pm {\bm J}^S$,
and Eq.~\ref{eq:ftg.gen} reduces to the fluctuation theorems derived earlier
(Eqs.~\ref{eq:ftg}, \ref{eq:ftg_dual}).
Further generalizations are relatively obvious --
for instance, replacing Eq.~\ref{eq:Lrev.gen} by
${\cal L}_\lambda^{\bm K} \leftrightarrow \{ {\bm K} , \varphi , {\bm G} \}$,
where ${\bm K}({\bm x};\lambda)$ is an arbitrary divergenceless vector field --
and will not be pursued here.

\section{Discussions and conclusions}

There has been considerable recent interest in understanding fluctuation theorems
within a Langevin
framework~\cite{03vZC,04RCWSES,04KQ,05Sei,05AA,05Kur,06Ast,06KQ,ImparatoPeliti06}.
This interest has been motivated in part by experiments involving systems that
are naturally modeled using Langevin dynamics, such as
externally manipulated biomolecules~\cite{02LDSTB,05CRJSTB},
optically trapped colloids~\cite{02WSMSE,04CRWSSE,04TJRCBL,05WRCWSSE,06BSHSB},
or a torsional pendulum~\cite{05DCPR,05DCP}.
In the present paper, using the path-integral framework,
we have emphasized
the separate contributions of two distinct sources of
nonequilibrium behavior: non-conservative forces and non-autonomous dynamics.

We conclude with a specific physical example where the exact relations discussed above apply.
Consider a macromolecule, such as a polymer, that is subject to thermal noise
and is manipulated externally (for instance using optical tweezers or atomic-force
microscopy) on a time scale comparable to its relaxation scale.
If such an experiment is carried out with the molecule immersed in a stationary,
equilibrium solution, then we have the situation of conservative forces
and non-autonomous dynamics.
As pointed out by Hummer and Szabo~\cite{01HuSz}
and subsequently verified in Refs.~\cite{02LDSTB,05CRJSTB},
such experiments can be used to reconstruct the free energy landscape
of the macromolecule.

We can imagine broadening the scope of these investigations to include non-conservative forces,
for instance by manipulating a molecule subject to shear~\cite{99SBC,00HSL}
and/or chaotic~\cite{00GS,04GS} flows.
If such flows are accurately modeled as Markovian noise --
that is, if the correlations of the flow decay on a time scale
that is very short in comparison with the relaxation time of the macromolecule --
then the results we have derived in this paper ought to apply directly.
When the shear or chaotic flows are non-Markovian,
then a suitable modification of our formalism would be needed.
A natural starting point for such a modification is the recent
progress achieved in the understanding of
polymer stretching and particle separation in chaotic flows \cite{00BFL,00Che}.

The feasibility of any realistic experiment along these lines will
be constrained by certain general considerations.
While the theoretical predictions are expressed in terms of
infinitely many realizations,
in a real experiment the number of measurements is necessarily finite.
At the same time, if the parameter $\lambda$ is varied rapidly,
then averages such as those in Eq.~\ref{eq:IFTs}
might be dominated by realizations that are extremely atypical.
The faster the protocol for varying $\lambda$, the more realizations
we are likely to need.
This introduces an important consideration when choosing a protocol
for externally manipulating the system.
Furthermore,
in experiments with complex molecules we inevitably have
access to only a small fraction of the system's degrees of
freedom, such as its end-to-end extension.
Thus we must design an experimental setup in which the quantities
that we want to measure ($Y$, $R$) can be determined uniquely
from the data to which we have access.
This might significantly limit the scope of the predictions that we could hope to test.

\begin{acknowledgments}
This research was supported by the Department of Energy, under
contract W-7405-ENG-36 and WSR program at LANL (M.C., C.J.),  and by
startup funds at WSU (V.Y.C.). C.J. thanks R.~Dean Astumian and Udo Seifert
for stimulating correspondence and discussions on this topic.

\end{acknowledgments}

\appendix
\section{Discretization Scheme}
\label{sec:Disc}

A stochastic differential equation such as Eq.~\ref{Lang1} remains
ambiguous until we specify the limiting procedure associated with the discretization
of time ($\Delta\rightarrow 0$).
In this Appendix we describe the discretization scheme that we adopt in this paper,
and we sketch the steps then taken to derive Eq.~\ref{eq:fp0}.
To avoid a proliferation of vector and tensor indices,
we restrict ourselves to the one-dimensional case.

We imagine that the time interval $[-\tau,\tau]$ is cut into $K\gg 1$ equal segments
of duration $\Delta$,
delimited by the sequence $\{t_0,t_1,\cdots,t_K\}$, where
$t_0=-\tau$, $t_K=\tau$, and $t_{k+1}-t_k=\Delta$.
The evolution of the dynamical variable $x(t)$ and external parameter $\lambda(t)$
are represented by a discrete sequence:
$(x(t),\lambda(t))\to (
x(t_k),\lambda(t_k))\equiv(x_k,\lambda_k)$.
The conditional probability of the trajectory $X = \{ x_0,\cdots,x_K \}$ is given by
a product of transition rates:
\begin{eqnarray}
 && {\cal P}_\lambda[X \vert x_0] = \prod\limits_k W_k(
 x_{k+1},x_{k}),\label{Pdisc1}\\
 \label{eq:midPoint}
 && W_k(x_{k+1},x_{k})= \tilde n(\overline{x}_k)
 \exp\left[-\frac{\Delta}{2}\left(\frac{\varepsilon_k}
 {\Delta}-F(\overline{x}_k)\right)^2\Gamma(\overline{x}_k)\right]
 \label{disc1}
\end{eqnarray}
where $\overline{x}_k\equiv(x_k+x_{k+1})/2$,
$\varepsilon_k \equiv x_{k+1}-x_k$,
and primes denote derivatives.
The prefactor $\tilde n$ is given by the function
\begin{equation}
\tilde n(x)=\sqrt{\frac{\Gamma(x)}{2\pi\Delta}}
 \left(1-\frac{\Delta}{\Gamma}\left[\frac{\Gamma'(x)^2}{4\Gamma(x)^2}-
 \frac{\Gamma''(x)}{8\Gamma(x)}\right]\right)
  \left(1-\frac{\Delta}{2} F'(x)\right) ,
\label{Ndisc}
\end{equation}
which guarantees normalization to first order in $\Delta$:
\begin{eqnarray}
\int dx_{k+1} W_k(x_{k+1},x_{k})=1 + O(\Delta^2). \label{Norm}
\end{eqnarray}
In the limit $\Delta\rightarrow 0$, Eqs.~\ref{Pdisc1} - \ref{Ndisc}
define the Markov process that we have denoted by Eq.~\ref{Lang1}.
Our task is now to derive the master equation associated with this Markov process.

Eq.~\ref{eq:midPoint} uses a mid-point regularization scheme
not only for the functions $F(x)$ and $\Gamma(x)$ inside the exponent,
but also for the prefactor $\tilde n(x)$.
Let us rewrite this prefactor,
\begin{equation}
\label{eq:ndef}
\tilde n(x) \equiv n(x)
\left(
1 - \frac{\Delta}{2} F^\prime
\right)
 \approx
 n(x)\,
 e^{-\Delta F^\prime/2},
\end{equation}
and incorporate the factor $e^{-\Delta F^\prime/2}$ inside the exponent
in Eq.~\ref{eq:midPoint}.
Thus, to first order in $\Delta$,
\begin{equation}
\label{eq:finalW}
W_k(x_{k+1},x_{k}) = n(\overline{x}_k)
 \exp\left[-\frac{\Delta}{2}\left(\frac{\varepsilon_k}
 {\Delta}-F(\overline{x}_k)\right)^2\Gamma(\overline{x}_k)
- \frac{\Delta}{2} F^\prime(\overline{x}_k) \right] ,
\end{equation}
with $n(x)$ as defined by Eqs.~\ref{Ndisc}, \ref{eq:ndef}.
This expression, combined with Eq.~\ref{Pdisc1}, specifies the sense in which
Eq.~\ref{eq:condProb.lambdaFixed} is interpreted, with
${\cal N} = \prod_{k=0}^{K-1} n(\overline{x}_k)$.
Note that ${\cal N}[X] = {\cal N}[X^\dagger]$ (because we use
the mid-point rule),
and ${\cal N}$ does not depend on the function $F(x)$.
These properties lead to the cancellation of normalization factors in
Eqs.~\ref{eq:ratio00} and \ref{eq:ratio.dual0}, respectively.

To arrive at Eq.~\ref{eq:fp0}, we must evaluate
\begin{equation}
 p_{k+1}(x) \equiv \int dx^\prime \, W_k(x,x^\prime) \, p_k(x^\prime)
 = \int d\varepsilon \, W_k(x,x-\varepsilon) \, p_k(x-\varepsilon),
 \label{pn+1}
\end{equation}
where $p_k$, $p_{k+1}$ denotes the probability distribution at time $t_k$, $t_{k+1}$.
Substituting Eq.~\ref{eq:finalW} into Eq.~\ref{pn+1},
and changing the variable of integration from $\varepsilon$ to
\begin{equation}
z \equiv \sqrt{\frac{\Gamma(x)}{\Delta}} \varepsilon,
\end{equation}
we get
\begin{equation}
\label{H}
p_{k+1}(x) =
\sqrt{\frac{1}{2\pi}}
\int dz \, \exp \left(-\frac{z^2}{2} \right) \, p_k(x) \, H(x,z,\sqrt{\Delta}) .
\end{equation}
We will not give the explicit, cumbersome expression for the function $H$
defined by this procedure, but we note that
$\lim_{\Delta\rightarrow 0} H(x,z,\sqrt{\Delta}) = 1$.
Thus we have factored out the dominant
Gaussian contribution to the integrand in Eq.~\ref{pn+1}, leaving a slowly varying
function of $z$.
Expanding $H$ in powers of $\sqrt\Delta$, to $O(\Delta)$,
and evaluating the resulting Gaussian integrals
(of the form $\int dz\, e^{-z^2/2} z^m$), we eventually obtain
\begin{equation}
 p_{k+1}=p_k-\Delta
 \left(
 F'p_k+
 Fp_k'
 +\frac{\Gamma'}{2\Gamma^2}
 p_k'-\frac{1}{2\Gamma}
 p_k''
 \right)  ,
\end{equation}
where the $p$'s, $F$, and $\Gamma$, and their derivatives,
are all evaluated at $x$.
In the limit $\Delta\to 0$, this becomes Eq.~\ref{eq:fp0}.

We emphasize that with a different choice of discretization -- for instance,
if we had used a prefactor that is a function of $x_k$ rather than the
mid-point $\overline{x}_k$ -- we would have obtained a different
Fokker-Planck equation.


\begin{thebibliography}{99}

\bibitem{93ECM} D.J. Evans, E.G.D. Cohen, G.P. Morris, Phys.Rev.Lett. {\bf 71}, 2401-2404 (1993).

\bibitem{94ES} D. Evans, D. Searles, Phys. Rev. E {\bf 50}, 1645 (1994).

\bibitem{95GC} G. Gallavotti, E.G.D. Cohen,
Phys.Rev.Lett. {\bf 74}, 2694-2697 (1995).

\bibitem{98Kur} J. Kurchan,
J.Phys.A {\bf 31}, 3719
(1998).

\bibitem{99LS} J.L. Lebowitz, H. Spohn,
J. Stat.Phys. {\bf 95}, 333 (1999).

\bibitem{77BoKu}
G.N.\ Bochkov and Yu.E.\ Kuzovlev,
Zh.Eksp.Teor.Fiz. {\bf 72}, 238 (1977)
[Sov.Phys.--JETP {\bf 45}, 125 (1977)].

\bibitem{97Jar1} C. Jarzynski,
Phys. Rev. Lett. {\bf 78}, 2690 (1997).

\bibitem{97Jar2} C. Jarzynski,
Phys.Rev. E {\bf 56}, 5018 (1997).

\bibitem{98Cro} G. E. Crooks, J. Stat. Phys. {\bf 90}, 1481 (1998).

\bibitem{99Cro} G.E. Crooks, Phys.Rev.E. {\bf 60}, 2721-2726 (1999).

\bibitem{99Hat} T. Hatano, Phys.Rev.E. {\bf 60}, R5017 (1999).

\bibitem{00Cro} G. E. Crooks, Phys.Rev. E {\bf 61}, 2361 (2000).

\bibitem{01HuSz} G. Hummer, A. Szabo, PNAS {\bf 98}, 3658 (2001)

\bibitem{01HS} T. Hatano, S. Sasa, Phys.Rev.Lett. {\bf 86}, 3463 (2001).

\bibitem{Mae03} C.~Maes, S\' em.~Poincar\' e {\bf 2}, 29 (2003).

\bibitem{MN03} C.~Maes and K.~Netocny, J.~Stat.~Phys.~{\bf 110}, 269 (2003).

\bibitem{04CCJ} V. Chernyak,  M. Chertkov, C. Jarzynski,
Phys.~Rev.~E {\bf 71}, 025102(R) (2005).

\bibitem{05Sei} U.~Seifert, Phys. Rev. Lett. {\bf 95}, 040602 (2005).

\bibitem{05Kur} J.~Kurchan, {\tt cond-mat/0511073}.

\bibitem{05RSE} J.C.~Reid, E.M.~Sevick, and D.J.~Evans, Europhys.\ Lett.\ {\bf 72}, 726 (2005).

\bibitem{05SpeckSeifert}
 T. Speck and U. Seifert, J. Phys. A: Math. Gen {\bf 38}, L581 (2005).

\bibitem{ImparatoPeliti06} A.~Imparato and L.~Peliti, {\tt cond-mat/0603506}.

\bibitem{02ES} D.J.\ Evans and D.\ Searles, Adv.\ Phys.\ {\bf 51}, 1529 (2002).

\bibitem{Rit03} F.~Ritort, S\' em.~Poincar\' e {\bf 2}, 193 (2003).

\bibitem{PS04} S.~Park and K.~Schulten, J.\ Chem.\ Phys.\ {\bf 120}, 5946 (2004).

\bibitem{05BLR} C.\ Bustamante, J.\ Liphardt, and F.\ Ritort, Physics Today {\bf 58}, 43 (2005).

\bibitem{06SpeckSeifert}
For a Brownian particle in one dimension,
analogues of our Eqs.~\ref{eq:JS} and \ref{eq:defA} are given by Eqs.~2 and 3
of T. Speck and U. Seifert, Europhys. Lett. {\bf 74}, 391 (2006).

\bibitem{Norris97}
J.~R.~Norris, {\it Markov Chains} (Cambridge University Press,
Cambridge, 1997).

\bibitem{Kemeny76}
J.~G.~Kemeny, J.~L.~Snell, and A.~W.~Knapp,
{\it Denumerable Markov Chains}, 2nd ed.\ (Springer-Verlag, New York, 1976).

\bibitem{non-generic}
Note that $p^S$ is unaffected by the angular drift force $f(q) \hat{\bm\theta}$.
This non-generic feature of our model arises from the fact that the flow
$f(q) \hat{\bm\theta}/\gamma$ is everywhere parallel to the contours of the harmonic potential $kq^2/2$.

\bibitem{53OM}
L.~Onsager and S.~Machlup,
Phys.\ Rev.\ {\bf 91}, 1505 (1953) and Phys.\ Rev.\ {\bf 91}, 1512 (1953).

\bibitem{wiegel}
F.W.~Wiegel,
{\it Introduction to Path-Integral Methods in Physics and Polymer Science}
(World Scientific, Philadelphia, 1986).

\bibitem{99BDKA}
M.~Bier, I.~Der\' enyi, M.~Kostur, and R.D.~Astumian,
Phys.\ Rev.\ E {\bf 59}, 6422 (1999).

\bibitem{98OP}
Y. Oono and M. Paniconi, Prog. Theor. Phys. Suppl. {\bf 130}, 29 (1998).

\bibitem{03vZC}
R.\ van Zon and E.G.D.\ Cohen,
Phys.\ Rev.\ E {\bf 67}, 046102 (2003); Phys.\ Rev.\ Lett.\ {\bf 91}, 110601 (2003).

\bibitem{04RCWSES}
J.C.~Reid, D.M.~Carberry, G.M.~Wang, E.M.~Sevick, D.J.~Evans, and D.J.~Searles,
Phys.\ Rev.\ E {\bf 70}, 016111 (2004).

\bibitem{04KQ}
K.H.~Kim and H.~Qian,
Phys.\ Rev.\ Lett.\ {\bf 93}, 120602 (2004).

\bibitem{05AA}
G.~Adjanor and M.~Ath\` enes,
J.\ Chem.\ Phys.\ {\bf 123}, 234104 (2005).

\bibitem{06Ast}
R.D.~Astumian, A.\ J.\ Phys., in press.

\bibitem{06KQ}
K.H.~Kim and H.~Qian, {\tt physics/0601085}.

\bibitem{02LDSTB} J. Liphardt, S. Dumont, S.B. Smith, I. Tinoco, C. Bustamante, Science {\bf 296}, 1832 (2002).

\bibitem{05CRJSTB}
D.~Collin, F.~Ritort, C.~Jarzynski, S.B.~Smith, I.~Tinoco, C.~Bustamante,
Nature {\bf 437}, 231 (2005).

\bibitem{02WSMSE}
G.M.~Wang, E.M.~Sevick, E.~Mittag, D.J.~Searles, and D.J.~Evans,
Phys.\ Rev.\ Lett.\ {\bf 89}, 050601 (2002).

\bibitem{04CRWSSE}
D.M.~Carberry, J.C.~Reid, G.M.~Wang, E.M.~Sevick, D.J.~Searles, and D.J.~Evans,
Phys.\ Rev.\ Lett.\ {\bf 92}, 140601 (2004).

\bibitem{04TJRCBL}
E.H.~Trepagnier, C.~Jarzynski, F.~Ritort, G.E.~Crooks, C.J.~Bustamante, and J.~Liphardt,
Proc.\ Natl.\ Acad.\ Sci.\ U.S.A.\ {\bf 101}, 15038 (2004).

\bibitem{05WRCWSSE}
G.M.~Wang, J.C.~Reid, D.M.~Carberry, D.R.~Williams, E.M.~Sevick, D.J.~Searles, and D.J.~Evans,
Phys.\ Rev.\ E {\bf 71}, 046141 (2005).

\bibitem{06BSHSB}
V.~Blickle, T.~Speck, L.~Helden, U.~Seifert, and C.~Bechinger,
Phys.\ Rev.\ Lett.\ {\bf 96}, 070603 (2006).

\bibitem{05DCPR}
F.~Douarche, S.~Ciliberto, A.~Petrosyan, and I.~Rabbiosi,
Europhys.\ Lett.\ {\bf 70}, 593 (2005).

\bibitem{05DCP}
F.~Douarche, S.~Ciliberto, and A.~Petrosyan,
J.\ Stat.\ Mech.: Theor.\ Exp.\ P09011 (2005).

\bibitem{99SBC}
 D. E. Smith, H. P. Babcock, and S. Chu, Science {\bf 283}, 1724 (1999).

\bibitem{00HSL}
 J. S. Hur, E. Shaqfeh, R. G. Larson, J. Rheol. {\bf 44}(4), pp. 713-741 (2000).

\bibitem{00GS}
 A. Groisman and V. Steinberg ,
 Nature {\bf 405}, 53 (2000);
 Phys. Rev. Lett. {\bf 86}, 934 (2001).

\bibitem{04GS}
 A. Groisman and V. Steinberg, New J. Phys. {\bf 6}, 29 (2004).

\bibitem{00BFL} E. Balkovsky, A. Fouxon, V. Lebedev,
Phys. Rev. Lett. {\bf 84}, 4765 (2000);
 Phys. Rev. E {\bf 64}, 056301 (2001).

 \bibitem{00Che}
 M. Chertkov, Phys. Rev. Lett. {\bf 84}, 4761 (2000).

\end{thebibliography}
\end{document}